\documentclass{emulateapj}

\usepackage{graphicx} 
\usepackage{color} 
 
\newcommand{\be}{  \begin{eqnarray} }
\newcommand{\ee}{  \end{eqnarray} }

\def\msun{{\rm M}_\odot}

\def\spose#1{\hbox to 0pt{#1\hss}}
\def\lta{\mathrel{\spose{\lower 3pt\hbox{$\mathchar"218$}}
     \raise 2.0pt\hbox{$\mathchar"13C$}}}
\def\gta{\mathrel{\spose{\lower 3pt\hbox{$\mathchar"218$}}
     \raise 2.0pt\hbox{$\mathchar"13E$}}}

\begin{document}

\shorttitle{Two-Phased ULX}
\title{Ultraluminous X-ray Sources Powered by Radiatively Efficient 
Two-Phased Super-Eddington Accretion onto stellar mass Black holes}
\author{Aristotle Socrates\altaffilmark{1,2} and Shane W. 
Davis\altaffilmark{3}}
\altaffiltext{1}{Department of Astrophysical Sciences, Princeton 
University, Peyton Hall-Ivy Lane, Princeton, NJ 08544; 
socrates@astro.princeton.edu}
\altaffiltext{2}{Hubble Fellow}
\altaffiltext{3}{Department of Physics, University of California, 
Santa Barbara, CA 93106; swd@physics.ucsb.edu}
\begin{abstract}

The radiation spectra of many of the brightest ultraluminous X-ray
sources (ULXs) are dominated by a hard power law component, likely
powered by a hot, optically thin corona that Comptonizes soft seed
photons emitted from a cool, optically thick black hole accretion
disk.  Before its dissipation and subsequent conversion into coronal
photon power, the randomized gravitational binding energy responsible
for powering ULX phenomena must separate from the mass of its origin
by a means other than, and quicker than, electron scattering-mediated
radiative diffusion.  Therefore, the release of accretion power in
ULXs is not necessarily subject to Eddington-limited photon trapping,
as long as it occurs in a corona.
  
Motivated by these basic considerations, we present a model of ULXs
powered by geometrically thin accretion onto stellar mass black holes.
In the region closest to the hole (region I), where the majority of
the binding energy is released, cool thermal disk radiation is
Comptonized by an adjacent corona covering the entire surface of the
disk.  The amount of reprocessed thermal emission in region I is quite
small compared to the hard coronal output since the disk behaves as a
near perfect reflector of X-rays, a result of the intense ionizing
flux which leads to an extreme state of photo-ionization.  If energy
injection takes place within an optically thin corona, the conversion
of binding energy into a wind is hampered by Compton drag and the
wind's low optical depth.  Furthermore, if the magnetic field geometry
of the corona is primarily closed, then magnetic fields of modest
strength can in principle, prevent the launching of a wind.  In the
outer regions (region II), where the albedo is somewhat lower, thermal
emission resulting from a combination of viscous dissipation within
the body of the disk and reprocessed coronal power is emitted at
relatively low temperatures, due to the large surface area.  Within
the context of the current black hole X-ray binary paradigm, our ULX
model may be viewed as an extension of the very high state observed in
Galactic sources.

\end{abstract}

\keywords{accretion, accretion disks -- black hole physics -- X-rays:
binaries -- X-rays: galaxies}

\section{Overview of Observations, current interpretations,
 and plan of this work}

Per detected photon, ultraluminous X-ray sources (ULXs) garner a
disproportionate amount of attention.  The reason for such relative
popularity results from the two most common interpretations of their
observed behavior: 1) super-Eddington luminosities resulting from
accretion onto stellar mass black holes 2) sub-Eddington accretion
onto intermediate mass black holes.  Both explanations are problematic
from a theoretical point of view.  With respect to super-Eddington
luminosities, it is not clear how a radiatively efficient flow can
release binding energy at a rate above the Eddington limit in a
steady-state manner.  As for intermediate mass black holes, a
well-established evolutionary path that accounts for their birth
does not exist.

Below, we summarize some important observational features of ULXs and 
previous theoretical models of their behavior in hopes to motivate our
own work.    

\subsection{Observations}\label{ss:observations}

We are primarily concerned with ULXs which display isotropic X-ray
luminosities $L_X\gtrsim 10^{40}{\rm erg\,s^{-1}}$ -- of order 10
times the Eddington limit for a 10$M_{\odot}$ black hole.  The
radiation spectra of these sources measured by {\it Chandra} and {\it
XMM} can be roughly fit by a power law with spectral index $\Gamma\sim
2$ (see Miller \& Colbert 2004 for a comprehensive review).  For
highly studied ULX sources, fits to the data allow for the presence of
a soft thermal component with $kT\sim 0.1-0.5\,{\rm keV}$ in addition
to a $\Gamma\sim 2$ power law.  The presence of such an anomalously
cool thermal component in many ULXs, along with the large inferred
X-ray luminosity, provides the best argument for the presence of
intermediate mass black holes.  That is, large luminosities emitted
thermally at relatively low temperatures necessarily requires a
relatively large emitting area and therefore, more massive black
holes.

\begin{figure}[b]
\includegraphics[width=0.46\textwidth]{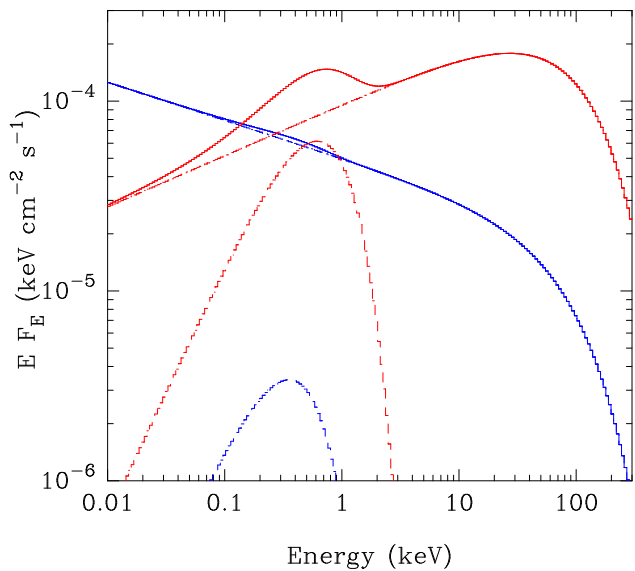}
\caption{The best fit MCD + power-law model spectra are plotted for
NGC 4038/4039 X-44 (blue) and M89 X-9 (2002 Apr 10; red).  The curves
represent the flux in the unabsorbed total model (solid), MCD
component (dashed), and power-law component (dot-dashed) for the
parameters listed in Table 1.  Note that the models were fit with a
power-law component over the 0.3-10 keV bands, but we have
extrapolated the models assuming 100 keV cut-off to the power-law for
the purpose of illustration.  We have also rescaled the overall flux
of M89 X-9 for illustrative purposes as well.\label{f:spectra}}
\end{figure}

Table 1 lists the salient observational features of highly studied
ULXs, selected on the basis of their large luminosities as well as the
ability of other workers to fit the data, in a statistically
meaningfully way, with a two-component model consisting of a power law
and soft multi-color disk black body.  Clearly, the power law
component in each individual source is a crucial component of the
total X-ray power.  To illustrate this point, Figure \ref{f:spectra}
provides best-fit theoretical spectra for two sources listed in Table
1.  By extrapolating the power law component to 100 keV, the potential
domination of the power law spectral component is further emphasized.

\begin{deluxetable*}{lccccccl}\label{t:sources}
\tablecolumns{8}
\tablecaption{Spectral Properties of Selected ULXs\label{tbl1}}
\tablewidth{0pt}
\tablehead{
\colhead{Source} &
\colhead{Date} &
\colhead{$k_{\rm B} T_d$} &
\colhead{$\Gamma$} &
\colhead{$F_{\rm pl}/F_{\rm total}$} &
\colhead{$L_X$} &
\colhead{$N_{\rm H}$} &
\colhead{Reference}\\
 &
 &
(keV) &
 &
 &
\colhead{$(10^{40}$ erg $s^{-1})$} &
\colhead{$(10^{21}\,{\rm cm}^{-2})$} &
}
\startdata
NGC 1313 X-1&
2000 Oct 17 &
$0.23 \pm 0.02$ &
$1.76 \pm 0.07$ &
$0.74$\tablenotemark{a}&
$0.6 \pm 0.1$\tablenotemark{a} &
$3.1 \pm 0.3$ &
 1\\
NGC 1313 X-2&
2000 Oct 17 &
$0.16^{+0.16}_{-0.04}$ &
$2.3^{+0.2}_{-0.1}$ &
$0.63$\tablenotemark{b} &
$0.66^{+0.18}_{-0.20}$\tablenotemark{b} &
$3^{+3}_{-1}$ &
 2\\
M81 X-9 &
2002 Apr 10&
$0.26^{+0.02}_{-0.05}$ &
$1.73 \pm 0.08$ &
$0.85$\tablenotemark{a} &
$1.1^{+0.3}_{-0.1}$\tablenotemark{a} &
$2.3 \pm 0.3$ &
1\\
M81 X-9 &
2002 Apr 16&
$0.21 \pm 0.04$ &
$1.86 \pm 0.06$ &
$0.84$\tablenotemark{a} &
$1.3^{+0.3}_{-0.2}$\tablenotemark{a} &
$2.9 \pm 0.3$ &
1\\
NGC 4038/4039 X-11 &
2002 Jan 8 &
$0.15 \pm 0.02$ &
$1.9 \pm 0.2$ &
$0.61$\tablenotemark{a} &
$2.1^{+0.7}_{-1.1}$\tablenotemark{a} &
$3.0^{+0.8}_{-1.8}$ &
3\\
NGC 4038/4039 X-16 &
2002 Jan 8 &
$0.19 \pm 0.05$ &
$1.4 \pm 0.2$ &
$0.33$\tablenotemark{a} &
$1.6^{+0.4}_{-1.0}$\tablenotemark{a} &
$1.5 \pm 1.0$ &
3\\
NGC 4038/4039 X-44 &
2002 Jan 8 &
$0.15^{+0.02}_{-0.15}$ &
$2.2^{+0.1}_{-0.4}$ &
$0.77$\tablenotemark{a} &
$1.0^{+1.3}_{-0.2}$\tablenotemark{a} &
$1.4^{+2.0}_{-0.4}$ &
3\\
NGC 4559 X-7 &
2003 May 27 &
$0.148 \pm 0.006$ &
$2.23^{+0.05}_{-0.04}$ &
$0.63$\tablenotemark{c}&
$2.2$\tablenotemark{c}&
$5.1^{+1.4}_{-1.3}$&
4\\
NGC 4559 X-7 &
2001 Jun 4 &
$0.12 \pm 0.006$ &
$1.8 \pm 0.08$ &
$0.69$\tablenotemark{c}&
$1.9$\tablenotemark{c}&
$3.6^{+0.9}_{-1.1}$&
4\\
NGC 4559 X-7 &
2002 Mar 14 &
$0.12 \pm 0.01$ &
$2.13 \pm 0.08$ &
$0.54$\tablenotemark{c}&
$3.2$\tablenotemark{c}&
$5.7^{+0.9}_{-1.1}$&
4\\
M74 X-1 &
2001 Oct 19 &
$0.31 \pm 0.05$ &
$1.83^{+0.55}_{-0.58}$\tablenotemark{e} &
$0.78$\tablenotemark{a}&
$0.4$\tablenotemark{a}&
$0.48$ &
5\\
Holmberg II X-1 &
2002 Apr 10 &
$0.141^{+0.018}_{-0.015}$ &
$2.64^{+0.06}_{-0.06}$ &
$0.83$\tablenotemark{d}&
$2.0$\tablenotemark{a}&
$1.6^{+0.2}_{-0.1}$ &
6\\
Holmberg II X-1 &
2002 Apr 16 &
$0.128^{+0.022}_{-0.013}$ &
$2.40^{+0.07}_{-0.08}$ &
$0.74$\tablenotemark{d}&
$1.7$\tablenotemark{a}&
$1.4 \pm 0.3$ &
6\\
Holmberg II X-1 &
2002 Sep 18 &
$0.120^{+0.022}_{-0.017}$ &
$2.89^{+0.07}_{-0.08}$ &
$0.68$\tablenotemark{d}&
$0.5$\tablenotemark{a}&
$1.4^{+0.5}_{-0.4}$ &
6\\
Holmberg II X-1 &
2004 Apr 15 &
$0.20 \pm {+0.02}$ &
$2.64 \pm 0.03$ &
$0.91$\tablenotemark{d}&
$1.2$\tablenotemark{d}&
$1.66^{+0.10}_{-0.09}$ &
7\\
NGC 5204 X-1 &
2003 Jan 6 &
$0.21 \pm {+0.03}$ &
$1.97 \pm 0.07$ &
$0.82$\tablenotemark{a,}\tablenotemark{f}&
$0.44 $\tablenotemark{a}&
$0.78^{+0.27}_{-0.17}$ &
8\\
NGC 5204 X-1 &
2003 Apr 25 &
$0.21^{+0.04}_{-0.03}$ &
$2.09^{+0.12}_{-0.13}$ &
$0.69$\tablenotemark{a,}\tablenotemark{f}&
$0.55 $\tablenotemark{a}&
$1.22^{+0.34}_{-0.29}$ &
8\\
M101 XMM-1 (P13/H19) &
2002 Jun 4 &
$0.30^{+0.04}_{-0.06}$ &
$1.41^{+0.25}_{-0.18}$ &
$0.74$\tablenotemark{a}&
$0.27 $\tablenotemark{a}&
$0.67^{+0.41}_{-0.19}$ &
9\\
NGC 7771 X-2  &
2002 June 21 &
$0.16^{+0.12}_{-0.03}$ &
$1.67^{+0.27}_{-0.26}$ &
$0.80$\tablenotemark{a}&
$3.86^{+0.1}_{-3.21} $\tablenotemark{a}&
$3.18^{+4.49}_{-2.51}$ &
10\\
\enddata
\tablenotetext{a}{Calculated over the energy range 0.3-10 keV.}
\tablenotetext{b}{Calculated over the energy range 0.2-10 keV.}
\tablenotetext{c}{Calculated over the energy range 0.3-12 keV.}
\tablenotetext{d}{Calculated over the energy range 0.3-2 keV.}
\tablenotetext{e}{Their definition of $\Gamma$ is equal to our
definition minus one}
\tablenotetext{f}{We calculated the ratio based on parameters
of their best-fit model.}
\tablerefs{(1) Miller et al. (2004a); (2) Miller et al. (2003); 
(3) Miller et al. (2004b) ; (4) Cropper et al. 2004; (5) Krauss et al.
2005; (6) Dewangan et al. 2004; (7) Goad et al. 2005; (8) Roberts 
al. 2005; (9) Jenkins et al. 2004; (10) Jenkins et al. 2005}

\end{deluxetable*}    

\subsection{Previous Interpretations}

Several models have been put forth to explain ULX phenomenon. For
$L_X\gtrsim10^{40}$ erg $\rm s^{-1}$, accretion onto intermediate mass
black holes (IMBHs) receives much attention.  If interpreted as
emission from an optically thick accretion disk extending to the
inner-most stable circular orbit of the black hole, the temperature of
the soft thermal component inferred from the spectra of these sources
are consistent with black hole masses of $\sim10^2 - 10^3 \msun$.
However, the large fraction of non-thermal emission in these sources
(see Table 1) brings into question the robustness of this
interpretation.  That is, the IMBH argument is straightforward only
when the cool thermal component is associated with the majority of the
bolometric luminosity, rather than a {\it sub-dominant} fraction of
the gravitational power.

Alternatively, ULXs may result from accretion onto $\sim 10 \msun$
black holes.  In this case, the emission must either be beamed so that
the intrinsic energy release is less than that inferred (King 2001) or
the Eddington limit must be exceeded.  Begelman (2002) suggests that
the non-linear development of the photon bubble instability (Gammie
1998; Blaes \& Socrates 2003; Turner et al. 2005) might produce
channels in the accretion flow, allowing the disk luminosity to exceed
the Eddington limit since radiative trapping is overcome.  In this
work we propose an alternative scenario for generating steady state
super-Eddington luminosities from stellar mass black holes, with the
intent of reproducing the hard, non-thermal emission that dominates
the spectra of many bright ULXs.

\subsection{Plan of this work}

In the light of the rough observational guidelines outlined in
\S\ref{ss:observations} for ULXs, an accretion model describing ULX
behavior must contain certain indispensable ingredients. Most
importantly, the observed large fluxes should be released
non-explosively.  A good candidate is radiatively efficient accretion
onto a compact object, whose birth and origin are well understood.
Such an accretion flow must account for the large apparent coronal
output while simultaneously accounting for the sub-dominant and
relatively cool 0.1-0.5 keV thermal emission component.  

In the next section we provide the framework in which the above
observational constrains are satisfied.  In \S\ref{ss:trapping} the
central idea of this work is motivated and argued.  That is, the
release of accretion power resulting from a two-phased accretion flow
is not bound by the Eddington limit since radiative diffusion,
mediated by electron scattering, is not primarily responsible for the
removal of randomized binding energy from the mass of its origin.  The
feasibility of confining coronal magnetic fields as well as the
effects of Compton drag on the outgoing optically thin wind are
discussed in \S\ref{ss:magnetic} and \S\ref{ss:wind}, respectively.
The formation of ULX spectra in terms of a corona+disk model is
considered in \S\ref{s:corona} and we conclude in \S\ref{conclusion}.

\section{the basic idea and initial estimates}\label{s:idea}

Perhaps the most important assumption of one-zone models of accretion
is that the randomized gravitational binding energy is dissipated
locally with the mass of its origin.  In the case of thin accretion
disks, the gravitational power is transported vertically via radiative
diffusion from the midplane of disk, where most of the mass resides.
As a result, super-Eddington accretion rates do not proportionally
lead to super-Eddington luminosities since the radial inflow time
becomes short relative to the time it takes for a photon to escape
from the disk midplane.  That is, the photon trapping radius moves
outward with accretion rate, negating the increase in available
binding energy with a decrease in the overall radiative efficiency.

Another consequence of the local dissipation assumption is that the
emergent radiation spectrum will be thermal and approach the black
body limit.  However, spectral components that probe the central
engines of active galactic nuclei (AGN) and X-ray binaries (XRBs)
indicate the presence of powerful non-thermal coronal emission
accompanied by a soft quasi-thermal component.  A common
interpretation of AGN and XRB spectra is two-phased accretion onto a
black hole (Haardt \& Maraschi 1991, 1993, hereafter HM91, HM93;
Svennsson \& Zdziarski 1994, hereafter SZ94).  In these models, a
fraction $f$ of the gravitational energy release is assumed to
dissipate in a hot diffuse corona, above the main body of the disk,
away from the majority of the accreted mass.  It is typically thought
that the viscous stress, assumed to be magnetic in nature (Balbus \&
Hawley 1991, hereafter BH91), transports angular momentum and
initially randomizes the gravitational binding energy near the
midplane.  The magnetized fluid elements, which are buoyant with
respect to their surroundings, dissipate -- perhaps through magnetic
reconnection or shocks -- above the disk where the reconnection time
and Mach number are large.
        
For large values of $f$, often necessary to fit XRB and AGN data, the
non-local dissipation of binding energy can have profound consequences
on the disk structure as well as the net radiative efficiency.  By
introducing a mechanism of energy transfer other than radiative
diffusion, the relevance of the Eddington limit is diminished (but not
eliminated entirely).  Below, we qualitatively discuss why
super-Eddington luminosities may result for two-phased accretion onto
a black hole.

\subsection{Altered disk structure and trapping radius}
\label{ss:trapping}

Following SZ94, we assume that all of the accretion takes place in the
main body of the disk.  Since we are interested in explaining ULX
behavior with stellar mass black holes, we restrict our discussion to
accretion rates above the Eddington limit for a 3-20 $M_{\odot}$ black
hole, implying a radiation pressure dominated disk.

The difference in structure between a standard thin disk and that of a
disk + corona is due the condition of radiative equilibrium \be
F_d=\sigma T^4_{eff}&=&\left(1-f\right)Q=\left(1-f\right)
\frac{3}{8\pi}\frac{GM{\dot M}}{R^3}\frac{{\mathcal D}} {{\mathcal B}}
\ee where $Q$ is the viscous dissipation rate per unit area.  All
changes to disk quantities due to coronal dissipation depends on
$\left(1-f\right)$ and for reference, their values (as well as
definitions of the various disk quantities) with respect to standard
thin disk scalings are given in Appendix \ref{a:diskquant}.  Most
notably, when compared to its classic one-zone counterpart, the disk
scale height $H_d$ decreases by a factor $(1-f)$ and therefore, the
cool disk remains thin as long as an increase in $\dot{m}$ is
proportionately matched by an increase in $f$.  Furthermore,
decreasing the internal dissipation increases the disk midplane
density increases by a factor $(1-f)^{-3}$.  This has the important
consequence that now, the radial inflow velocity becomes 
\be
v_R=\frac{{\dot M}}{4\pi\,RH_d\rho_d}=v_{0R}\left(1-f\right)^2 \propto
{\dot M}^2\left(1-f\right)^2 
\ee 
where $v_{0R}$ is the classic thin
disk value (SS73).  As long as increasing $\dot{m}$ is matched by an
increase in $f$, then $v_R$ remains fixed, implying a reduction in the
level of radiative trapping.  To quantify this statement even further,
we define the trapping radius $R_{tr}$, defined as the radius at which
the radial inflow time is equal to the vertical diffusion time, 
\be
R_{tr}\sim \frac{1}{2^{1/2}}R_g\left(1-f\right)^{1/2}\dot{m}. 
\ee 
As expected, an increase in $\dot{m}$ is matched by an increase in
$R_{tr}$, directly leading to a decrease in radiative efficiency.  For
$f\sim 0$ at the Eddington accretion rate, the trapping radius
$R_{tr}\sim 10 R_g$ for a Schwarzschild hole.  This implies that thin
disk accretion cannot maintain its radiative efficiency beyond the
Eddington accretion rate if the release of binding energy is governed
by electron scattering-mediated radiative diffusion.  However, for
large values of $f$ the binding energy of the gas is removed from
large to small optical depths by a mechanism other than radiative
diffusion, perhaps magnetic buoyancy or waves.  In this case, the
modified radial inflow time $R/v_R$ must be compared to the
characteristic timescale of say, buoyant transport.

Without a coherent theory of relativistic accretion disk turbulence,
it is difficult to estimate what the characteristic speed of
magnetized buoyant transport $v_B$ will be.  A good guess for the
upper limit of $v_B$ is the sound speed of the disk at the midplane
$c_s$.  Let us define an inward advection time $\tau_{adv}\equiv
R/v_R$ and a timescale for upward buoyant transport $\tau_B\equiv
H_d/c_s$.  As long as the ratio $\tau_{adv}/\tau_B > 1$, the accretion
power of the flow stands a chance of escaping before being consumed by
the black hole.  In terms of disk quantities, we have
\be
\frac{\tau_{adv}}{\tau_B}\lesssim \frac{r^{1/2}\tau_d}{\dot{m}}
\simeq \frac{r^2}{\alpha\,\dot{m}^2\left(1-f\right)^2}
\label{e:trap}
\ee 
If we (somewhat arbitrarily) take $\alpha$ to be $\sim 0.1$, then the
above ratio assumes a minimum value of $\tau_{adv}/\tau_B \sim 5$ at
$r\sim 12$, the radius of maximum light, for an accretion rate rate of
$ {\dot m}=175$ -- ten times the Eddington rate -- for a choice of
$(1-f)\sim 0.1$. Therefore, the efficiency of the disk remains
unchanged as long as an increase in the accretion rate $\dot{m}$ is
proportionately matched by a decrease in $(1-f)$.  Even if the rate at
which randomized gravitational binding energy leaks upwards decreases
by a factor of $\sim 5$, the disk may maintain its radiative efficiency
for accretion rates of order $\sim 10$ times the Eddington rate.

\subsection{Magnetic field requirements}\label{ss:magnetic}

Any viable accretion model that steadily generates super-Eddington
luminosities must fulfill two requirements.  First, the accretion flow
must remove its randomized binding energy before being advected into
the hole in order to maintain a large radiative efficiency.
Furthermore, the majority of that binding energy must be converted
into radiation rather than a mechanical outflow.  The central point of
this paper, encapsulated by eq. (\ref{e:trap}), addresses the first
requirement.  That is, magnetic buoyancy is an ideal mechanism for the
non-dissipative removal of binding energy from the bulk of the flow.
Addressing the second requirement is conceptually more difficult and
again, we appeal to the existence of magnetic fields.

Irrespective of the geometry of the emitting surface, a
super-Eddington radiation flux produces an outward opposing
acceleration that is larger than the attractive inward acceleration
provided by gravity (see e.g., Abramowicz et al.  1980).  Therefore,
an additional force or stress is required in order for our
super-Eddington accretion model to maintain its radiative efficiency.
Begelman (2002) suggests that strong magnetic fields might prevent
surface regions of low optical depth from being blown off by a
super-Eddington flux generated near the disk midplane.  Furthermore,
several authors have noted that accretion flows which liberate a
sizable fraction of their binding energy in a Comptonzing corona
inevitably require modest magnetic fields in order to mitigate the
escape of pairs by the act of confinement (Svensson 1984; Zdziarski
1985; White \& Lightman 1989). We examine a similar scenario in the
context of the super-Eddington flux arising from energy injection in
an optically thin corona. If indeed the confinement is provided for by
magnetic fields, then there are some straightforward constraints
regarding its topology and strength.


Charged particles are free to move along magnetic field lines.  As
long as the field lines are anchored in the disk, the supposed site of
their generation, their magnetic tension is capable of confining fluid
elements.  Numerical simulations of stratified accretion flows
indicate that the coronal field geomtry is primarily closed and
toroidal (Miller \& Stone 2000; Hirose et al. 2004).  However, it is
not clear whether or not field lines are anchored deep in the disk. To
gain an order of magnitude estimate of the acceleration a confining
magnetic field can impart to a electrically conducting coronal fluid
element, we define $g_M$ to be a characteristic magnetic acceleration
as
\be
g_{M}\sim \frac{v^2_A}{H_c}\sim 10^{13}
\left(\frac{v_A}{0.1 c}\right)^{2}
\left(\frac{R_g}{H_c}\right)\left(\frac{10}{m}\right)\,{\rm cm\,s^{-2}},
\label{e:g_m}
\ee
which should be compared to the magnitude of the gravitational 
acceleration
\be
g\sim 10^{15}\left(\frac{10}{m}\right)\left(\frac{H_c}{R_g}\right)
r^{-3}\,\frac{{\mathcal C}}{{\mathcal B}}\,{\rm cm\,s^{-2}},
\ee
and the radiative acceleration
\be
g_{rad}=\frac{\kappa_{es}}{c}F_d \sim 2\times10^{17}\left(
\frac{10}{m}\right)\left(\frac{\dot m}{175}\right)r^{-3}\frac{\mathcal D}
{\mathcal B},
\label{e:g_rad}
\ee
which has a maximum value of $g_{rad}\sim 2\times 10^{13}$ for ${\dot
m}=175$ at $r\sim 12$.  Note, $\dot{m}$ is the accretion rate scaled
to the Eddington rate $\dot{M}_{Edd}=4 \pi G M m_p/(c \sigma_T)$ and
$m$ is the black hole mass in units of $M_{\odot}$. If the radiative
force in the corona is balanced by gradients in the magnetic field,
then the ratio $g_M/g_{rad}\geq 1$ and the quantity $g_{rad}/g$ is
approximately the factor by which the Eddington limit is surpassed.
If we assume that $H_c$ and $R$ scale in proportion to the mass of the
hole, then like $g$, the characteristic magnetic and radiative
acceleration $g_M$ and $g_{rad}$ are $\propto M^{-1}_{\rm BH}$ and
therefore in principle, similar Eddington ratios can be achieved for
supermassive black holes.

The magnetic field strength or pressure in the corona can be
constrained as well.  First, the magnetic pressure of the corona
$P_{m,c}\sim B^2_c/8\pi$ must be at least comparable to the coronal
pressure.  The other requirement is that the coronal magnetic pressure
can not exceed the magnetic pressure at the disk midplane.  Following
the scalings of SZ94, the required coronal field strength is given by
equating $P_{m,c}$ to $P_{rad,c}$ (see Appendix\ref{a:diskquant}),
which yields
\be
B_c \gtrsim 2.4 \times 10^8 \,
\dot{m}^{1/2}\Lambda^{1/2} r^{-3/2} \left(\frac{\mathcal{D}}{\mathcal{B}}
\right)^{1/2}\,{\rm G}
\label{e:B_c}
\ee
for $f\sim 1$.  By assuming a magnetic origin for the accretion
torque, an estimate of the magnetic field strength in the disk
midplane is obtained by setting $P_{m,d}\sim \tau_{R\phi}\sim
\alpha P_d$, giving us
\be 
B_d \simeq 1.6 \times 10^8 \,
\left(1-f\right)^{-1/2} r^{-3/4} \left(\frac{\mathcal{C}}
{\mathcal{A}}\right)^{1/2}\,{\rm G}.
\label{e:B_d}
\ee
For ${\dot m}=175$, $f=0.9$, and $\Lambda \sim 1$ at a radius of
$r\sim 12$ the ratio $B_d/B_c\sim 3$, which is its lowest value.  Thus
it seems possible that if only a fraction of the dynamo-generated flux
resides in the corona, the plasma there can be contained despite the
upward force exerted upon it by the super-Eddington radiation
field.

As a check of internal consistency, the coronal magnetic field given by
eq. (\ref{e:B_c}) must lead to an Alfv\`en speed $v_A$ large enough to
satisfy eq. (\ref{e:g_m}).  In order to evaluate $v_A$, knowledge of
the coronal density is required.  For a radiation pressure dominated
corona, a characteristic density is roughly given in Appendix
\ref{a:diskquant}, leading to an
Alfv\'en speed of $v_A\sim c$ for the choice of disk and coronal
parameters previously mentioned.

\subsubsection{Hierarchy of stresses: Or, why coronae are essential 
for super-Eddington luminosities}

Consider the following relevant thought experiment.  Imagine
injecting thermal energy into the Sun at a rate of $\sim 10^{39}\,
{\rm erg\,s^{-1}}$, which is $\sim 10$ times its Eddington rate.
Regardless of the depth at which this enormous deposition of energy
takes place, the local pressure will similarly increase by a factor
of $\sim 10$.  This can be understood by noting that the ratio of
radiation to gas pressure $P_{rad}/P_g$ is roughly a constant
throughout the star and approximately equal to its Eddington factor
such that $P_{rad}/P_g\sim L_{\odot}/L_{Edd}$.

If the injection takes place uniformly at a great depth near the core,
then the resulting radiation stress is $\sim 10$ times larger than any
other stress in the entire star.  Thus, the bulk of the star has no
choice other than to move outward and form a massive wind that carries
away the injected energy.  However, if the energy injection uniformly
takes place at the surface and near the photosphere, then the
situation can potentially, be quite different.  That is, a radiation
pressure supplying a flux $\sim 10$ times the Eddington limit at the
surface of the Sun is only $\sim 10^{-12}$ that of the gas pressure in
the core.  Hence, super-Eddington energy injection at the surface does
not appreciably change the structure of the star.  If the stresses
within the star re-arrange themselves only slightly, then a
super-Eddington flux due to energy injection at the surface can easily
be confined as to stifle the production of a non-radiative
super-Eddington wind.  To further clarify this point, take the example
of a sunspot, whose magnetic stresses are supposedly generated by
highly subsonic (and highly sub-virial) turbulent and shearing motions
at the base of the solar convection zone.  For typical sunspot field
strengths $\sim$ a few $\times 10^3$ G, closed field lines at the
photosphere could in principle, confine matter that is accelerated by
radiation pressure of order $\sim 10$ times Eddington.  Despite the
fact that only an infinitesimal amount of the Sun's internal energy
resides in its convection zone, subsonic motions in these regions may
lead to stresses large enough to contain regions of the photosphere
experiencing a super-Eddington radiation pressure.

The situation described in the previous section with respect to ULXs
powered by super-Eddington accretion disks is quite similar to the
solar thought experiment outlined above.  Even for super-Eddington
luminosities, the ratio of coronal to midplane pressure $P_c/P_d$ is
quite small.  In the disk case, the magnetic accretion stress $\alpha
P_d$ originates from motions that are subsonic in the midplane by a
different mechanism than, but analogous to, the generation of field
loops in the shear layer at the base of the solar convection zone.
The arguments surrounding eqs. (\ref{e:B_c}) and (\ref{e:B_d})
indicate that if even a small fraction of the magnetic energy leaks
into photosphere, magnetic confinement is a plausible mechanism
for the suppression of a super-Eddington wind.

\subsection{Compton-driven wind}\label{ss:wind}

In the absence of closed magnetic field lines, it is commonly thought
that the majority of the gravitational power for a super-Eddington
accretion flow either is advected into the black hole or is converted
into mechanical energy that drives an outflow.  In this two-phased
accretion model, it is unlikely that a significant amount of the
binding energy is advected into the hole for sufficiently large
values of $f$, the fraction of power released in the corona.
Therefore, most of the radiative luminosity would be expected to be
converted into the kinetic power of the outflow.  However, this
conclusion follows from the assumption that the majority of the
energy injection takes place beneath the photosphere.

The properties of continuum driven winds follow from conservation of
momentum (Lamers \& Cassineli 1999)
\be
\label{e:windmom}
\rho v\frac{d v}{d R}+\frac{d P_g}{d R}+\rho \frac{G M}{R^2}=
\frac{n_p\sigma_T }{c} \frac{L}{4\pi R^2} \left(1+2\frac{n_+}{n_p}\right)
\ee
and mass
\be
\dot{M}=4\pi R^2 \rho v.
\ee 
Here, $n_p$ is the number density of the ions and $n_+$ is the number 
density of pairs. For the sake of simplicity, we have assumed spherical
symmetry.  Since we are considering super-Eddington luminosities, the
forces due to gravity and gas pressure can be neglected outside the
sonic point.  Integrating over space, eq. (\ref{e:windmom}) then becomes
\be
\dot{M} v_{\infty} \simeq \tau_p\frac{L }{c}
\left(1+2\frac{n_+}{n_p}\right),
\ee
where $\tau_p$ and $v_{\infty}$ are the plasma optical depth and outflow
velocity at infinity, respectively.  Note that the ratio $n_+/n_p$ is
approximated as constant.  The above expressions are accurate for 
non-relativistic velocities.  Due to the large radiative acceleration,
a relativistic outflow seems inevitable.   However, Compton 
drag limits the Lorentz factor $\Gamma_L$ to moderate values, 
such that $\Gamma_L\gtrsim 1$ (Madau \& Thompson 2000). 

For $v_{\infty}\sim c$, the kinetic luminosity in the wind $L_w$
is simply
\be
L_w\sim \tau_p\,L\left(1+2\frac{n_+}{n_p}\right).
\ee 
Thus, the kinetic power resulting from super-Eddington energy 
injection is directly proportional to the total plasma +
pair scattering optical depth and in principle, may be significantly
smaller than the photon luminosity.  These arguments are simple.
In order to truly quantify the rate of loss of mechanical energy, 
the spatial distribution of energy injection must be specified, 
potentially allowing for a detailed calculation of the coronal 
structure, pair equilibrium, and resulting spectrum.

\begin{figure*}[t]
\input{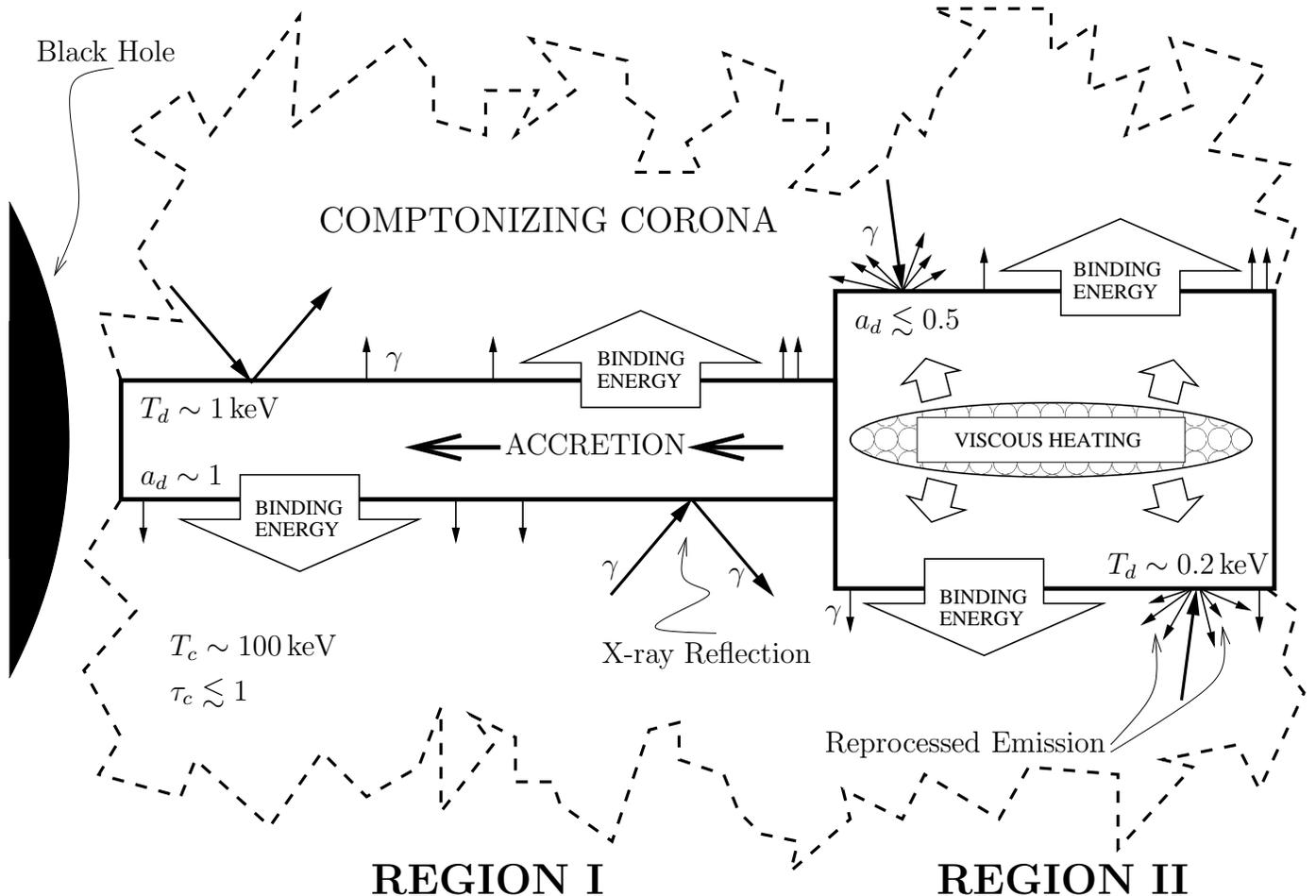}
\caption{Sketch of our stellar mass black hole ULX accretion model.
In both Regions I and II, gravitational binding energy is removed from
the disk in the vertical direction before it is converted into heat.
However, accretion power may dissipate locally in Region II without
being advected into the hole.  The majority of the gravitational power
is released in Region I primarily in the form of hard Comptonized
photons.  Throughout the flow, cool photons from the optically thick
disk are Comptonized in the surrounding hot diffuse corona.  Hard
X-ray photons incident upon the disk are nearly perfectly reflected in
Region I while in Region II, they are photo-electrically absorbed by
the disk.  In the corona, the production of an energetically dominant
super-Eddington wind is hindered by the presence of confining magnetic
fields, a low optical depth, and the effects of Compton drag.  Note
that this figure is not to scale.\label{f:model}}

\label{f:model}
\end{figure*}

\section{Coronal Structure and ULX Spectra}\label{s:corona}

The radiation spectra of high luminosity ULXs are often dominated by a
relatively flat $\Gamma\sim 2$ power law.  Also, the fractional
thermal emission, responsible for $\sim\frac{1}{3}$ to $\frac{1}{10}$
of the luminosity, is typically modeled by a black body or
multi-temperature disk black body with a characteristic temperature of
0.1-0.5 keV.  Along with producing a luminosity of $\sim
10^{40}-10^{41}{\rm erg\,s^{-1}}$, these two observational constraints
must be reproduced by an accretion flow spectral model that describes
ULX behavior.  

Our model consists of a two-phased accretion flow divided into two
distinct geometrically thin regions (see Figure \ref{f:model}).  The
inner portion of the disk, or region I, is responsible for liberating
the majority of the system's gravitational binding energy, while the
outer portion, region II, powers the sub-dominant thermal spectral
component.

In region I, we adopt the slab geometry for the disk+corona i.e., the
simplest configuration possible.  In order to reproduce the spectra of
luminous ULXs, region I cannot be a prodigious site of soft thermal
photon power.  Therefore, the emergent spectra from the region of deepest
gravitational potential must resemble a $\Gamma\sim 2$ Comptonized 
power law with only a small fraction of the luminosity in the form 
of thermal emission such that ratio of hard coronal to soft power
$L_c/L_s\sim 10$ in region I.  

As described in the previous section, if randomized binding energy is
deposited in a corona in excess of the Eddington rate, hydrostatic
balance cannot be achieved unless forces that are non-gravitational in
nature, such as magnetic stresses, are present.  In the absence of
magnetic fields, the corona becomes geometrically thick and a wind
develops.  Calculating the structure of the corona by accurately
including the relevant processes that accompany magnetic field
generation and transport, non-equilibrium radiation physics, and winds
is an impossibly difficult problem, with no clear way as to how to
proceed. However, some simplification is achieved by realizing that
Comptonization of ambient soft photons must be the radiative mechanism
which mediates the release of energy in the corona.  That is,
relatively few physical parameters, such as optical depth and
temperature, are required to reproduce the observed spectral
components for a given coronal geometry.  Though such an approach 
lacks almost any sense of predictive
power, it provides a rough outline as to what the flow geometry and
energetics might be.

In what follows, we briefly discuss the composite physical 
requirements of the accretion flow necessary to reproduce the 
observed spectra.

\subsection{Spectral index, coronal temperature, and optical depth
of the inner region} 

An almost universal feature of the spectral energy distributions
(SEDs) of the relativistic accretion flows that power active galactic
nuclei (AGN) and XRBs is a predominantly flat broadband X-ray power law
which results from thermal Comptonization of hot coronal electrons.
In AGN and black hole XRBs in the low/hard state, the spectral index
$\Gamma\lesssim 2$, with greater variation seen in black hole XRBs.

In the case of low/hard state spectra of galactic black hole
candidates such as Cyg X-1, both the local and disk-integrated values
of $L_c/L_s\gg 1$.  The most popular disk+corona geometry that is
invoked to explain these sources is a truncated disk + hot inner
spherical corona model (Poutanen, Krolik, \& Ryde 1997; Esin et
al. 1998).  The inner hot regions are thought to resemble the
ADAF/CDAF accretion models that are thought spontaneously occur at low
accretion rates (Narayan \& Yi 1995; Quataert \& Gruzinov 1999).
However, there are other explanations for the low/hard state that
cannot be ruled out and both involve altering the patchy corona model
of HMG93 by reducing the soft reprocessed disk emission.  If the
overwhelming majority of the flow's gravitational power is liberated
in relatively small active coronal regions, then the ionizing flux
incident upon the disk will be so large that the atomic X-ray
absorptivity of the disk becomes vanishingly small and disk acts
almost as a perfect reflector (Ross \& Fabian 1993; Ross, Fabian, \&
Young 1999; Ballantyne, Ross, \& Fabian 2001).  Another mechanism that
reduces reprocessed disk emission in the patchy corona model is
outward vertical motion of the active regions at sub-relativistic
velocities (Beloborodov 1999).

In region I of our ULX model, the local and disk-integrated value of
$L_c/L_s\gg1 $ as in the case of black hole XRBs in the low/hard
state.  For a spectral index $\Gamma\sim 2$, equal photon power is
emitted at all photon energies bounded by the input seed photon energy
and the coronal electron temperature.  Therefore, the majority of the
disk photons interact with the corona only a few times, if any, and
the relatively few photons that stay in the corona the longest carry
away with them most of the radiated power.  The likelihood for a
photon to escape the corona, which varies inversely with the optical
depth $\tau$, is appropriately balanced by the average photon energy
shift per scattering $\Delta E/E\sim 4 k_BT_c /m_ec^2$ in order to
reproduce a given spectral index $\Gamma$ (see e.g., Fermi 1949 within
the context of cosmic ray acceleration in the interstellar medium).
Thus, if the $k_BT_e\sim 100$ keV, then only $\frac{1}{100}$ of the
input photons would have to leave the system at that energy for a soft
seed photon energy of $\sim 1$ keV and an outgoing spectral index
$\Gamma\sim 2$.  Such harmony can be achieved for $\tau\sim 1$ since
$\Delta E/E\lesssim 1$ if $k_BT_e\sim 100$ keV.  That is, a
$\Gamma\sim 2$ Comptonized spectrum is viable only when the propensity
of escape per scatter is proportionally countered by a gain in energy
for every scattering event.

For flat $\Gamma \sim 2$ Comptonized spectra, the amplification
$L_c/L_s$ can be estimated by calculating the number of scattering
orders $N_s$ required to fill out the Comptonized power law, since the
location of each scattering order in the SED carries equal amounts of
energy as the soft input. By choosing soft seed photon energies of
order $\sim 1$ keV, roughly an Eddington's worth of power is emitted
from the cold optically thick disk.  From the argument above, $\Gamma
\sim 2$ for coronal temperatures of order $\sim 100$ keV and an optical
depth $\sim 1$.  For these parameters, we obtain $N_s\sim
L_c/L_s\lesssim 10$.  Thus, an input soft seed X-ray source typical
for optically thick Eddington-limited accretion onto 10$\msun$ black
hole is amplified up to $\sim 10$ times its intrinsic thermal power if
there is an adjoining corona with $T_e\sim 100$ keV and $\tau\sim 1$
covering the entire disk.

The somewhat prosaic arguments given above are in rough agreement with
detailed Comptonization calculations of other authors (Pietrini and
Krolik 1995; Stern et al. 1996).  For example, Pietrini and Krolik
(1995) deduced a scaling relationship between $T_e$, $\tau$,
$L_c/L_s$, and $\Gamma$ that reproduce the desired spectral properties
of our ULX coronal model.  Similar to the low/hard state of black hole
XRBs, our ULX model requires a low level of soft photon input.  We
cannot appeal to a truncated disk + ADAF/CDAF scenario because large
accretion rates and high radiative efficiencies are necessary for
ULXs.  In order to reduce the amount of reprocessed thermal power, the
albedo of the disk must approach unity due to the intense
super-Eddington ionizing flux.  In what follows, we discuss the
validity of this assertion.

\subsection{Albedo}       

In order to reproduce spectra dominated by a Comptonized power law
in the slab geometry, the Comptonized photons incident upon the 
disk must avoid photo-electric absorption.  If a significant
fraction of the hard X-ray flux is absorbed by the disk, the reprocessed
thermal emission can in principle be comparable to the local X-ray 
flux such that $L_c/L_s\sim 1$ in the slab geometry
(Haardt \& Maraschi 1991).  Thus, either the disk albedo must approach 
unity as a result of photo-ionization (Ross \& Fabian 1993) or the
coronal electrons flow away from the disk at 
sub-relativistic to relativistic velocities (Beloborodov 1999).       

In a super-Eddington model of ULXs, the ionizing coronal flux is 
intense in comparison to say, ionizing disk models of black hole
XRBs in the low/hard state.  An increase in disk albedo implies 
a change in the disk's role from that of an absorber and re-emitter
to a reflecting X-ray mirror.  If the disk albedo $a_d$ approaches its 
limiting value such that $a_d=1$, nearly 
all of the metal ions in surface of the disk must be fully ionized.   

In Appendix \ref{a:albedo} $a_d$ is shown to closely approach unity
near the inner-edge of the disk for ${\dot m}= 175$ while sharply
dropping off at larger radii.  By $r\sim 60$, outside of which $1/5$
of the gravitational power is released, $a_d\sim 1/2$.  Therefore, one
expects a significant amount, at least compared to the Eddington
limit, of thermal emission to be emitted from ULX flows, despite the
majority of the flow's gravitational power being released via
Comptonization.

Eq. (\ref{e:epsilon}) raises an interesting point.  For accretion
disks that power AGN, the abundant CNO metals are not fully ionized
since the thermal disk temperatures are significantly lower.  As a
result, $A_Z\sim 10^{-2}-10^{-3}$, implying that the disk cannot act
as a perfect reflector and the spectra will be different than the XRB
case in that the reprocessed thermal spectral component is more likely
to be comparable in magnitude to the Comptonized power law.

\subsection{Thermal component}

The inferred thermal component of bright ULXs are often cited as the
most compelling evidence for the existence of intermediate mass black
holes due to their relatively low temperatures.  Roughly speaking, the
most luminous sources can be fit just as well with a simple power law
rather than a sub-dominant disk + dominant power law component.
Therefore, explaining the physical processes responsible for the
inferred soft thermal component is not central to our understanding
of ULXs.  Nevertheless, we describe the manner in which soft thermal 
emission may emerge from our ULX model.

In our model, thermal disk emission primarily originates from region
II.  Take the example of a $10{\msun}$ black hole accretion flow which
resembles our ULX model accreting at $10$ times the Eddington rate
${\dot m}_{Edd}$.  For a non-rotating hole, the brightest region of
the flow is located at $r\sim 12$ and since gravitational power falls
of as $\propto r^{-1}$, roughly $~90\%$ of the accretion power is
liberated within $r\sim 120$.  Assume that this luminous inner portion
of the flow constitutes region I.  Outside this radius in region II,
further assume that the dissipation and release of gravitational
binding energy proceeds in the classical SS73 sense such that an
Eddington's worth of gravitational power is emitted there as a result
of viscous dissipation.  In this case, the disk remains thin at $r\sim
120$ such that $H/R\lesssim 1$, the diffusion time is short compared
to the inflow time, and the radiation spectra is roughly thermal.
Compared to a disk with ${\dot m}={\dot m}_{Edd}$ around the same
black hole, the temperature of the thermal emission from region II in
the ULX case is cooler due to its larger emitting surface.  Typical
inner-disk effective temperatures for an Eddington accretor around a
$10\msun$ black hole are $k_BT_{eff}\sim 0.7\,{\rm keV}$, whereas the
${\dot m}=10\,{\dot m}_{Edd}$ ULX model radiates an equal amount of
power at a lower temperature of $\sim 0.2-0.3\,{\rm keV}$.

An energetically dominant corona is required in order to generate more
than an Eddington's worth of thermal power from region II in order to
avoid the trapping and thickness problem.  Therefore, super-Eddington
values of thermal power must result from the reprocessing of coronal
photons inside the trapping radius.  If we increase the accretion rate
to $50\,{\dot m}_{Edd}$, then for the same emitting area for region II
discussed above, $\sim 5$ times an Eddington's worth of thermal power
will be emitted at a temperature of $\sim 0.4-0.5\,{\rm keV}$ outside
of $r\sim 120$.  Evidently, our ULX model possesses the ability to
produce over an Eddington's worth of thermal power for a $10\msun$
black hole at relatively low temperatures in what we have termed
region II.

To mitigate the radiative inefficiencies due to the production of a
powerful wind, which inevitably accompanies super-Eddington thermal
fluxes, region II's corona must possess and ordered, predominantly
azimuthal magnetic field.  By use of eqs. (\ref{e:B_c}) and
(\ref{e:B_d}), we see that the required coronal field strength $B_c$
in region II is small compared to value of the field in the midplane
such that $B_c/B_d\lesssim \frac{1}{10}$ for sufficiently large values of
$f$.


\section{Discussion and Summary}\label{conclusion}

Many bright ULXs exhibit power-law emission with hard $\Gamma\lesssim
2$ spectral indexes.  In fact, recent {\it Chandra} survey data
indicates that ULXs on average have $\Gamma\simeq 1.8$ with
significant scatter, but independent of luminosity (Swartz et al.
2004).  If ULXs are simply scaled up versions of Galactic black hole
XRBs, then their $\Gamma\lesssim 2$ power-laws are consistent with the
low/hard state spectra of Galactic sources (Remillard \& McClintock
2003).  For this scenario, the accretion flow is expected to be
sub-Eddington such that $L_X\sim 10^{-2}\, L_{edd}$ (e.g. Maccarone
2003), implying black hole masses $\gtrsim 10^4 M_{\odot}$.  In the
ADAF/CDAF model of the hard state, the thermal emission is produced by
a truncated disk at relatively large radii and correspondingly low
temperatures.  If we adopt this picture for ULXs, then the temperature
of the truncated disk component would be a factor of 3-7 times lower
than the $\sim 0.1$ keV truncated disk temperature found in Galactic
black hole XRBs such as Cyg X-1 (Gierlinski et al.  1997; Poutanen,
Krolik, \& Ryde 1997).  As a result, the thermal components near 
$\sim 0.1$ keV which are inferred from fits to some bright ULXs
would not be produced in the truncated disk model.

Thus, an explanation of the soft thermal component inferred from
bright ULX spectra requires a disk + corona geometry that is different
from the above picture for Galactic XRBs, whether or not the central
object is an IMBH.  If ULXs are indeed powered by an IMBH
ADAF/CDAF-type accretion flow, then the soft spectral component must
come from another source such as diffuse nebular emission.  In any
case, for hard $\Gamma\lesssim 2$ power-laws, the soft component of
ULXs cannot be utilized to deduce the mass of the black hole in a
straightforward manner, unless the overall accretion geometry is
markedly different from that of Galactic XRBs.

Our accretion model is primarily motivated by the fact that many
bright ULXs are dominated by hard non-thermal photon power.  This
strongly suggests that radiative diffusion is not responsible for
transporting the majority of the gravitational binding energy out of
the accretion flow.  Photon trapping, a mechanism which caps the
luminosity of standard thin/slim disks, need not limit the flow's
output to the Eddington value.  For super-Eddington energy injection
rates, the outward radiative acceleration outstrips the local
gravitational acceleration.  If reasonably strong magnetic stresses
are present in the corona, the generation of a continuum-driven wind
is prevented as long as the coronal field line are closed and anchored
in the disk.  In the absence of confining magnetic fields, the
resulting wind inefficiently converts photon energy into outflowing
mechanical power due to the effects of Compton drag and a low optical
depth.  Therefore, the majority of the super-Eddington accretion power
is not necessarily converted into mechanical energy and thus, the
flow's radiative efficiency is left relatively untouched.  In the
corona, randomized gravitational binding energy is converted into
radiative power via Compton amplification of soft seed photons.  As a
result of the large downward ionizing flux deep in the gravitational
potential, the disk's albedo is sufficiently high such that it behaves
as a reflecting X-ray mirror and relatively little thermal emission
emerges from the luminous inner regions.  Further away from the black
hole a combination of reprocessed emission and viscous dissipation may
produce roughly an Eddington's worth of photon power at low thermal
energies since the emitting area is comparatively large.

Phenomenologically, our ULX model may be viewed as a natural extension
of the commonly accepted picture of how Galactic black hole XRBs
evolve with luminosity (e.g. Done \& Gierli\'nski 2004).  For soft
state XRBs, the ratio of coronal to soft power $L_c/L_s$ increases
with luminosity.  In the ``high state'' $L_c/L_s\sim 0.1$ whereas in
the ``very high state,'' typical values of $L_c/L_s\sim 1$.
Furthermore, $L\sim 0.1\, L_{Edd}$ in the high state while the
luminosity of the very state approaches $L_{Edd}$ and interestingly,
$L_c/ L_s\sim 10$ for ULXs with $L\sim 10\,L_{Edd}$.
 
Of course, there are many theoretical uncertainties uniquely haunting
the ULX model presented in this work.  In order to overcome the
Eddington limit by a factor of $\simeq 10$, the fraction of binding
energy released in the corona $f\simeq 0.9$ so that the bulk of the
flow can expel its binding energy before being advected into the hole,
while remaining geometrically thin.  If magnetic fields are utilized
as a confining mechanism in the corona, their strength, geometry, and
stability must be considered in further detail.  The net radiative
efficiency of the flow hinges upon the assertion that the majority of
the accretion power does not escape in a mechanical form.  In order to
verify this claim, the vertical structure of the corona resulting
from the intense Comptonizing radiation field, must be be determined.

\appendix

\section{A: Disk Scalings}\label{a:diskquant}

In this appendix, we provide disk scalings relevant for the 
disk midplane following the work of SZ94.  When the accretion
rate is large we choose to ignore radial advection, an
assertion that is justified as long as the fraction of the 
energy dissipated in the corona $f\simeq 1$ i.e., the regime
relevant for ULX behavior.

In the radiation pressure dominated limit, the midplane 
disk quantities are given by
\be
\frac{H_d}{R}=\frac{3}{2}{\dot m}r^{-1}(1-f)\frac{{\mathcal D}}{{\mathcal 
C}},
\ee   
\be
\rho_d=\frac{8}{9}\frac{m_p}{\sigma_TR_g}\alpha^{-1}{\dot m}^{-2}r^{3/2}
(1-f)^{-3}\frac{{\mathcal C}^2{\mathcal B}}{{\mathcal D}^2{\mathcal A}},
\ee
\be
\tau_d=\frac{4}{3}\alpha^{-1}{\dot m}^{-1}r^{3/2}(1-f)^{-2}
\frac{{\mathcal C}{\mathcal B}}{{\mathcal D}{\mathcal A}},
\ee
\be
P_{rad, d}=P_d=
\frac{2}{3}\frac{m_pc^2}{\sigma_TR_g}\alpha^{-1}r^{-3/2}(1-f)^{-1}
\frac{{\mathcal C}}{{\mathcal A}}.
\ee
The factors ${\mathcal A}-{\mathcal D}$ enforce the no-torque inner
boundary condition and take into account the effects of general
relativity (Riffert and Herold 1995; Hubeny and Hubeny 1998).  Note
that the dimensionless cylindrical radius $r$ is scaled to the
gravitational, rather than Schwarzschild radius such that
$r=R/R_g=Rc^2/GM$.  The four physical disk quantities above are
determined by four conservation laws.  Namely, conservation of mass,
vertical momentum (hydrostatic balance), angular momentum, and energy
(radiative equilibrium).

That is, we assume that it is possible for
all of the angular momentum transport and accretion to take place in
the body of the disk even if the overwhelming majority of the
dissipation resulting from the action of accretion takes place in the
corona.  Thus, in order to constrain these quantities in the
corona, we arbitrarily assume a value for the coronal optical depth
$\tau_c$.  In the radiation pressure dominated limit,
hydrostatic balance may be written as
\be
\frac{\sigma_T F_d}{m_p c}=\frac{G M}{R^2} \frac{z}{R} 
\frac{\mathcal C}{\mathcal B}
\ee
where the left hand side is the upward radiative acceleration associated 
with a 
flux $F_d$, and the right
hand side is the force of gravity in the thin disk limit.  Defining $H_c$ 
as the height above 
the midplane where $F_d=Q$ we find
\be
\frac{H_c}{R}=\frac{3}{2}\,{\dot m}r^{-1}
\frac{{\mathcal D}}{{\mathcal C}}.
\label{e:Hcorona}
\ee
Following SZ94, we estimate a coronal density from the scale height and 
optical depth
$\tau_c=H_c \rho_c \sigma_T/ m_p$.  

Inserting eq. \ref{e:Hcorona} for $H_c$ in the radiation pressure 
dominated case yields
\be
\rho_c=\frac{2}{3}\frac{m_p}{\sigma_TR_g}{\dot m}^{-1}
\,\tau_c\frac{{\mathcal C}}{{\mathcal D}},
\ee
For $\tau_c \lesssim 1$, the radiation pressure can be  approximated as
\be
P_{rad, c}\sim \Lambda \frac{Q}{c}=\frac{3 \Lambda}{2}\frac{c^2m_p}
{\sigma_TR_g}{\dot m}r^{-3}
\frac{{\mathcal D}}{{\mathcal B}}
\ee
where $\Lambda$ is a factor of order unity which is determined by the 
angular dependence of the radiation field.

\section{B: Estimate of Disk Albedo}\label{a:albedo}

The disk albedo $a_d$ reaches large values when $\epsilon$, the ratio
of absorption to total opacity, is small.  In order to
keep the photo-electric absorption rate at a low value, the disk must
be close to a fully ionized state, which is roughly determined by the
density of the gas, number of ionizing photons, and recombination
rate.  A comparison with the results of other works is facilitated by
the introduction of the ``ionization parameter'' $\xi_{I}$, given by
\be 
\xi_{I}\equiv \frac{4\pi F}{n_H} \simeq
\frac{4\pi\,m_pf\,Q}{\beta\,\rho_d} \simeq 2\times
10^6\left(\frac{0.1}{\beta}\right)\left(\frac{\alpha}{0.1}\right)
\left(\frac{{\dot m}}
{175}\right)^3\left(\frac{1-f}{0.9}\right)^3\left(\frac{f}{0.9}\right)
\left(\frac{r }{12}\right)^{-9/2}\,{\rm erg\, cm\,s^{-1}}.
\ee     
Here, $\beta$ parameterizes the density of the X-ray absorbing or
reflecting region near the surface of the disk and is always chosen
such that $\beta<1$.  We have eliminated the relevant relativistic and
no-torque correction factors, which take on a value $\sim 0.003$ at
$r\sim 12$.  Detailed spectral and ionization calculations show that
$a_d\sim 1$ when $\xi_{I}\sim 10^5$ (see Ballantyne 2002 for a concise
review).

In the following, we motivate why $a_d\sim 1$ for $\xi_{I} \sim 10^5$
erg cm s$^-1$.  For the sake of simplicity, we group all metals into a
single hydrogenic species with abundance $A_Z$.  This approximation is
only valid when the absorber/reflector is close to being fully ionized
and when the incident X-ray spectra is roughly described as having
equal power across photon energy since different ions have different
ionization potentials.  With this, the expression that governs the
balance between the number of hydrogenic and completely stripped ions
is given by 
\be
n_i\int_{\nu_0}^{\infty}\frac{F_{\nu}}{h\nu}\sigma\left(\nu\right)d\nu
=n_e\,n_{i+1}\alpha_R\left(T\right)
\label{e:ionization}
\ee        
where $n_i$, $n_{i+1}$, $n_e$, $F_{\nu}$, $\sigma\left(\nu\right)$,
and $\alpha_R\left(T\right)$ is the number density of bound hydrogenic
ions, number density of completely stripped ions, number density of
free electrons, incident flux, absorption cross section, and
recombination rate, respectively.  By assuming a flat $\Gamma=2$
spectral index for the ionizing continuum and that
$\sigma\left(\nu\right)=\sigma_0\left(\nu/\nu_0\right)^3$, an
expression for the hydrogenic fraction $\chi_i$ reads 
\be
\chi^{-1}_i\equiv
\frac{n_{i+1}}{n_i}=\frac{4\pi\,F_I}{n_H}\frac{\sigma_0}
{16\,h\,\nu_0\,\alpha_R\left(T\right)}\frac{n_H}{n_e}.  \ee Here
$F_I\equiv\int_{\nu_0}^{\infty} F_{\nu}$ and if we further approximate
that $F\sim F_I$ then by use of eq. (\ref{e:ionization}) \be
\chi^{-1}_i\sim 10^5\left(\frac{\xi_I}{2\times 10^6}\right),
\label{e:ionization}
\ee 
for typical values of the micro-physical parameters i.e.,
$\sigma_0\sim 10^{-18} \,{\rm cm^2}$, $h\nu_0\sim 10\,{\rm keV}$, and
$\alpha_R(T)\sim 10^{-10}\, {\rm s^{-1}}$.  Now, we are in a position
to calculate $\epsilon$ the ratio of absorption to scattering opacity
in terms of the fiducial metal abundance $A_Z$ 
\be
\epsilon\sim\frac{\kappa_{abs}}{\kappa_{sc}}\sim
\frac{\chi_i\,A_Z\,\sigma_0} {\sigma_T}\sim 5\times
10^{-4}\left(\frac{A_Z}{5\times 10^{-5}}\right) \left(\frac{2\times
10^6}{\xi}\right).
\label{e:epsilon}
\ee 
Note that the value of $A_Z$ corresponds to Fe at solar
abundances.  Due to the large disk temperatures in XRBs, the
relatively abundant CNO metals are fully ionized.  Finally, in order
to calculate the $a_d$, the disk albedo, we make use of the two-stream
approximation -- an accurate approximation in the elastic limit for
photon energies under $10$ keV -- for the disk surface layers (see
e.g. HM93).
\be a_d\simeq
\frac{1-\epsilon^{1/2}}{1+\epsilon^{1/2}}\simeq \frac{1-0.024}
{1+0.024}\simeq 0.953 
\ee 
for our choice of parameters at $r\sim 12$.

\acknowledgements{We thank O. Blaes, R. Kulsrud, B. Pacy\'nski,
J. Ostriker, E. Ramirez-Ruiz, J.Stone and D. Uzdensky for helpful
conversations.  SWD thanks the Institute for Advanced Study, where a
portion of this work was completed, for its hospitality. AS
acknolwdges support of a Hubble Fellowship administered by the Space
Telescope Science Institute.}

{}

\end{document}